# DNS Study on Vorticity Structures in Late Flow Transition

Xiangrui Dong[1, 2], Shuling Tian[3, 2], Chaoqun Liu[2, *]

[1]National Key Laboratory of Transient Physics, Nanjing University of Science & Technology, Nanjing, Jiangsu, 210094, China

[2] Department of Mathematics, University of Texas at Arlington, Arlington, Texas, 76019, USA

[3] College of Aerospace Engineering, Nanjing University of Aeronautics and Astronautics, Nanjing, Jiangsu, 210016, China

Vorticity and vortex are two different but related concepts. This paper focuses on the investigation of vorticity generation and development, and vorticity structure inside/ outside the vortex. Vortex is a region where the vorticity overtakes deformation. Vortex cannot be directly represented by the vorticity. Except for those vorticity lines which come from and end at side boundaries, another type of vorticity, self-closed vorticity lines named *vorticity rings*, is numerously generated inside the domain during flow transition. These new vorticity rings are found around the hairpin vortex heads and legs. The generation and growth of vorticity rings are produced by the buildup of the vortices according to the vorticity transport equation. On the other hand, vortex buildup is a consequence of vorticity line stretching, tilting and twisting. Both new vorticity and new vortices are generated during the flow transition. According to the Helmholtz vorticity flux conservation law, vorticity line cannot be interrupted, started, or ended inside the flow field, the newly produced vorticity has only one form which is the vorticity rings. In addition, an interesting finding is that a single hairpin vortex consists of several types of vorticity lines which could come from the side boundaries, whole vorticity rings and part of vorticity rings.

## Nomenclature

| | | | |
|---|---|---|---|
| $M_\infty$ | = Mach number | $Re_{in}$ | = Reynolds number |
| $\delta_{in}$ | = inflow displacement thickness | $U_\infty$ | = freestream velocity |
| $u, v, w$ | = components of velocity | $\rho$ | = density |
| $p$ | = pressure | $T$ | = temperature |
| $e$ | = total energy per mass unit | $k$ | = thermal conductivity |
| $\mu$ | = dynamic viscosity | $R$ | = ideal gas constant |
| $Pr$ | = Prandtl number | $\gamma$ | = ratio of specific heats |
| $C_P$ | = specific heat at constant pressure | $C_V$ | = specific heat at constant volume |
| $z^+$ | = first grid interval in the normal direction | $\vec{B}$ | = sum of the external body forces |
| $x, y, z$ | = streamwise, spanwise, normal directions | T | = a period of T-S wave |
| $\sigma_{xx}, \sigma_{yy}, \sigma_{zz}, \sigma_{xy}, \sigma_{xz}, \sigma_{yz}$ | = components of viscous stress | $Lz_{in}$ | = height at inflow boundary |
| $Lx$ | = length of computational domain along the $x$ direction | | |
| $Ly$ | = length of computational domain along the $y$ direction | | |

* Corresponding author. Email address: cliu@uta.edu

$x_{in}$   = distance between leading edge of flat plate and upstream boundary of computational domain

**Subscript**
*in*   = inflow
*w*   = wall
∞    = free stream

# 1 Introduction

Vorticity and vortex in fluid dynamics have been used interchangeably by many authors [1-3] in past decades. However, a distinction must be made between vortices and vorticity [4]. Vorticity $\boldsymbol{\omega} = \nabla \times \boldsymbol{u}$, where $\boldsymbol{u}$ is fluid velocity, has a rigorous mathematical definition – it is the curl of a velocity field [5]. Some concepts such as vorticity lines, vorticity tubes, vortex filaments, vortex (vortices) often confuse readers. In mathematics, a vorticity line is defined as a curve which is tangent to the local vorticity vector everywhere; the vorticity tube is a tube-like surface formed by all vorticity lines passing through a closed material curve, which could be considered as a bundle of vorticity lines. However, as denoted by Helmholtz [6], vortex lines are the lines drawn through the fluid mass so that their direction at every point coincides with the direction of the momentary axis of rotation of the fluid particles lying on it. Vortex filaments are also denoted by Helmholtz as the portions of the fluid mass cut out from it by way of constructing corresponding vortex lines through all points circumference of an infinitely small surface element. From Helmholtz, there is no doubt that a vortex is a fluid region with rotational motion. More explanations of vortex like 'the sinews and muscles of the fluid' and 'the sinews of turbulence' were respectively given by Kuchemann [7] and Moffatt et al. [8]. Several authors have proposed quantified definitions. Brachet et al. [9] defined a vortex as a region of negative velocity gradient determinant. For a long time, the vortex was tracked by vorticity or even directly regarded as the region with higher vorticity magnitude. Babiano et al. [10] defined a vortex as any region of a fluid with vorticity magnitude greater than some threshold. Wu et al. [1] presented in his book that the name "vortex tube" is imprecise, since the rigorous definition of a vortex is still a controversial issue, and the side boundary of a vortex is not a vorticity surface. He also claimed a vortex as a connected fluid region with relatively high concentration of vorticity. Although Green [2] gave a similar definition: a fluid vortex is any region of concentrated vorticity, which means the vorticity can be a measure of how rapidly fluid rotates about itself, he pointed out two exceptions, the laminar boundary layer and the sum of two Gaussian 'blobs' of vorticity. However, these definitions are inaccurate in some cases. In 1989, Robinson [11, 12] pointed out that the association between regions of strong vorticity and actual vortices can be rather weak in the turbulent boundary layer, especially in the near wall region. Similar conclusions were obtained by Wang et al. [13] – a vortex is not necessarily the congregation of vorticity lines, but a dispersion in most three dimensional cases, which means the vorticity in a vortex is not necessarily larger than the surrounding area in many 3D cases.

Several vortex identification criteria, such as the $\tilde{\Delta}$ method, $\lambda_2$, Q, and Omega (Ω) criteria [14-17], have been proved to be efficient in visualizing vortex structures when applied to DNS data of a transitional flow [18]. Although both vorticity and circulation well understand concepts, the objective definition of a vortex is still a difficult issue [19]. In addition, as Kolar [20] point out, the compressibility should be considered while most vortex identification methods assume the follow is incompressible by setting the first invariant P = 0. However, in our DNS, we pick the Mach number as 0.5, which has minor affection by the compressibility. Being different from many other DNS papers which mostly discuss vortex structure, this paper is focused more on the generation, development and structure of the new vorticity in late flow transition, and does more analysis on the relationship between vorticity and vortex and further investigates the vorticity types and distributions inside a vortex.

For deeper investigation into the comparison between vorticity structures and vortex structures (vortical structures), a boundary layer transitional flow on a flat plate at a freestream Mach number of 0.5 is utilized by a high order DNS. The vortex visualization of this study is performed by a so-called Omega (Ω) criterion [17] which has some advantages such as normalized from 0 to 1, fixed-threshold and capability to capture both strong and weak vortices. However, the authors have no intention to compare those vortex identification methods or make comments on these methods in this paper, but just use Omega (Ω) criterion as a tool for visualization of the vortex structures in flow transition. This paper is organized as follows: Section 2 presents the numerical methods and the case description; Section 3 provides our DNS results and addresses the details about the generation process of the vorticity structures, especially vorticity ring generation in the flow transition. The last is our conclusion.

## 2 Numerical Algorithms and Case Description

### 2.1 Governing Equations

The flow field is governed by a non-dimensional compressible Navier-Stokes system which can be written in the following conservative form:

$$\frac{\partial Q}{\partial t} + \frac{\partial F}{\partial x} + \frac{\partial G}{\partial y} + \frac{\partial H}{\partial z} = \frac{1}{Re}\left(\frac{\partial F_v}{\partial x} + \frac{\partial G_v}{\partial y} + \frac{\partial H_v}{\partial z}\right) \tag{1}$$

where the vector of conserved quantities $Q$, inviscid flux vector $E, F$ and $G$, and viscous flux vector $E_v, F_v$ and $G_v$ are

$$Q = \begin{pmatrix} \rho \\ \rho u \\ \rho v \\ \rho w \\ e \end{pmatrix}, F = \begin{pmatrix} \rho u \\ \rho u^2 + p \\ \rho uv \\ \rho uw \\ (e+p)u \end{pmatrix}, G = \begin{pmatrix} \rho v \\ \rho uv \\ \rho v^2 + p \\ \rho vw \\ (e+p)v \end{pmatrix}, H = \begin{pmatrix} \rho w \\ \rho uw \\ \rho vw \\ \rho w^2 + p \\ (e+p)w \end{pmatrix}$$

$$F_v = \begin{pmatrix} 0 \\ \sigma_{xx} \\ \sigma_{xy} \\ \sigma_{xz} \\ (u\sigma_{xx} + v\sigma_{xy} + w\sigma_{xz}) + \frac{1}{(\gamma-1)PrM_\infty^2}k(T)\frac{\partial T}{\partial x} \end{pmatrix}, G_v = \begin{pmatrix} 0 \\ \sigma_{xy} \\ \sigma_{yy} \\ \sigma_{yz} \\ (u\sigma_{xy} + v\sigma_{yy} + w\sigma_{yz}) + \frac{1}{(\gamma-1)PrM_\infty^2}k(T)\frac{\partial T}{\partial y} \end{pmatrix},$$

$$H_v = \begin{pmatrix} 0 \\ \sigma_{xz} \\ \sigma_{yz} \\ \sigma_{zz} \\ (u\sigma_{xz} + v\sigma_{yz} + w\sigma_{zz}) + \frac{1}{(\gamma-1)PrM_\infty^2}k(T)\frac{\partial T}{\partial z} \end{pmatrix}$$

The components of viscous stress are listed as follows,

$$\sigma_{xx} = \frac{2}{3}\mu(T)\left(2\frac{\partial u}{\partial x} - \frac{\partial v}{\partial y} - \frac{\partial w}{\partial z}\right), \sigma_{yy} = \frac{2}{3}\mu(T)\left(-\frac{\partial u}{\partial x} + 2\frac{\partial v}{\partial y} - \frac{\partial w}{\partial z}\right), \sigma_{zz} = \frac{2}{3}\mu(T)\left(-\frac{\partial u}{\partial x} - \frac{\partial v}{\partial y} + 2\frac{\partial w}{\partial z}\right),$$

$$\sigma_{xy} = \mu(T)\left(\frac{\partial u}{\partial y} + \frac{\partial v}{\partial x}\right), \sigma_{xz} = \mu(T)\left(\frac{\partial u}{\partial z} + \frac{\partial w}{\partial x}\right), \sigma_{yz} = \mu(T)\left(\frac{\partial w}{\partial y} + \frac{\partial v}{\partial z}\right)$$

The state equation is given in Eq.2:

$$e = \frac{p}{\gamma - 1} + \frac{1}{2}\rho(u^2 + v^2 + w^2) \tag{2}$$

Since the governing equations are non-dimensional, the reference values for length, density, velocity, temperature and pressure are $\delta_{in}, \rho_\infty, U_\infty, T_\infty$ and $\rho_\infty U_\infty^2$, respectively, where $\delta_{in}$ is the inflow displacement thickness. The Mach number $M_\infty$, Reynolds number $Re_{in}$, the ratio of specific heats $\gamma$ and the Prandtl number $Pr$ are expressed as,

$$M_\infty = \frac{U_\infty}{\sqrt{\gamma R T_\infty}}, \quad Re_{in} = \frac{\rho_\infty U_\infty \delta_{in}}{\mu_\infty}, \quad Pr = \frac{C_p \mu_\infty}{k_\infty}, \quad \gamma = \frac{C_P}{C_V}$$

Where $R$, $C_P$ and $C_V$ are the ideal gas constant and the specific heats at constant pressure and constant volume. $k_\infty$, $\mu_\infty$ is the thermal conductivity and the dynamic viscosity, respectively. Through this work, $Pr = 0.7$ and $\gamma = 1.4$.

## 2.2 Numerical Methods

A sixth order compact scheme [21] is adopted for the spatial discretization in the streamwise and normal directions, while in the spanwise direction where periodical conditions are applied, the pseudo-spectral method is used. Instead of artificial dissipation, a high-order spatial scheme filtering [21] is used to eliminate the spurious numerical oscillations caused by the central difference scheme. For time marching, a third order Total Variation Diminishing (TVD) Runge-Kutta scheme [22] is employed.

## 2.3 Case Description

The sketch of the physical domain of our DNS simulation is shown in Figure 1. *X*, *Y* and *Z* are dimensionless quantities in *x*, *y*, and *z* axes, and the reference values for length is the inflow displacement thickness $\delta_{in}$. The length of the computational domain in sreamwise and spanwise directions is *Lx* and *Ly*, and the height of the domain *Z* varies from $Lz_{in}$ by following the similarity $U = U(\eta), \eta = \frac{Z}{\sqrt{X}}$, where $Lz_{in}$ is the height of the domain at inflow boundary. $x_{in}$ is the distance between the inlet and leading edge of the flat plate. Detailed geometry parameters are listed in Table 1. The grid system [23] is $n_{streamwise} \times n_{spanwise} \times n_{normal}$ = 1920×128×241. The grid is stretched in the normal direction while uniform in both streamwise and spanwise directions. The first grid interval in the normal direction, $z^+$, is set as 0.43. The flow parameters are also listed in Table 1. The Reynolds number is defined as $Re_{in} = \rho_\infty U_\infty \delta_{in} / \mu_\infty$, here $\delta_{in}$ is the inflow displacement thickness, which acts as reference length. *Tw* and $T_\infty$ denote the wall temperature and the freestream temperature, respectively.

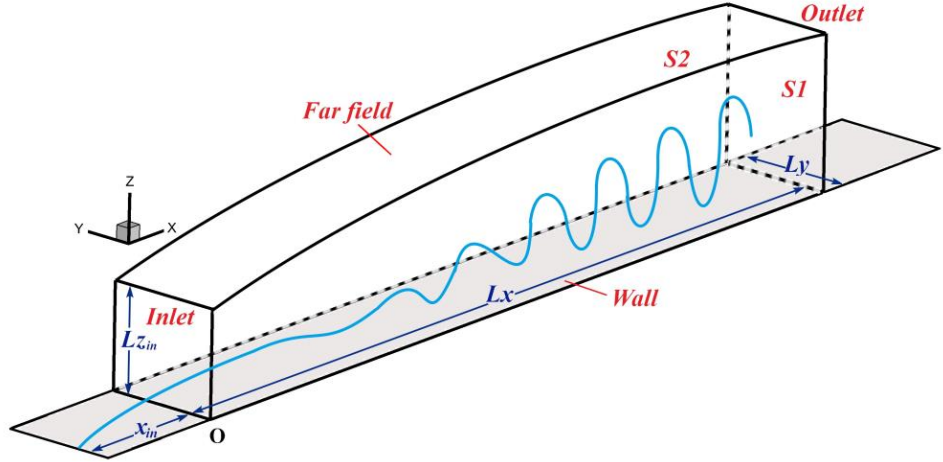

**Figure 1 Sketch of the flow transition domain**
**Table 1 Flow parameters & Geometry parameters**

| $M_\infty$ | $Re_{in}$ | $T_\infty$ | $T_w$ | $x_{in}$ | $Lx$ | $Ly$ | $Lz_{in}$ |
|---|---|---|---|---|---|---|---|
| 0.5 | 1000 | 273.15 K | 273.15 K | 300.79 $\delta_{in}$ | 798.03 $\delta_{in}$ | 22 $\delta_{in}$ | 40 $\delta_{in}$ |

## 2.4 Boundary Conditions

The inflow boundary conditions with perturbations can be given in the following equation,

$$q = q_{lam} + A_{2d} q'_{2d} e^{i(\alpha x - \omega t)} + A_{3d} q'_{3d} e^{i(\alpha x \pm \beta y - \omega t)} \qquad (3)$$

where *q* denotes *u*, *v*, *w*, *p*, and *T*, and $q_{lam}$ represents the Blasius solution for a 2D laminar boundary layer. The Tollmien-Schlichting (T-S) waves are obtained by solving the compressible boundary layer stability equations by picking $\omega = 0.114027$. The linear solution gives $\alpha = 0.29919 - i5.09586 \times 10^{-3}$, $\beta$ is set as zero in 2D T-S wave generation and set as 0.5712 to get the 3D T-S wave. We also specified $A_{2d} = 0.03$ and $A_{3d} = 0.01$ in our DNS. It should also be noted that the disturbances are enforced at the inflow computational boundary, which has a distance of $x_{in}$ downstream the leading edge of the plate (Figure 1). The top boundary is set as far field boundary condition and the outlet is treated as outflow boundary. Meanwhile, a good non-reflecting boundary condition [24] is applied on the far field boundary and the outflow boundary. The adiabatic and the non-slipping conditions are enforced at the wall boundary, while the periodic boundary conditions are utilized at the spanwise boundaries.

*2.5 Code Validation*

The DNSUTA code which was developed by University of Texas at Arlington is adopted in this study [25]. A short description of the validation will be addressed here. Figure 2 shows the skin friction coefficient, $C_f$, calculated from the time-averaged and spanwise-averaged profile in two different grid levels (coarse grids: 960×64×121, fine grids: 1920×128×241), the skin friction coefficient for the laminar flow and the turbulent boundary layer by Cousteix [26] is also given in the figure for comparison. The skin friction coefficient after $x \approx 450\delta_{in}$, which is defined as the 'onset point', is in a good agreement with the flat-plate theory of the turbulent boundary layer. This comparison indicates that a proper velocity profile for turbulent flow can be obtained from our DNS result, and the grid convergence has been verified.

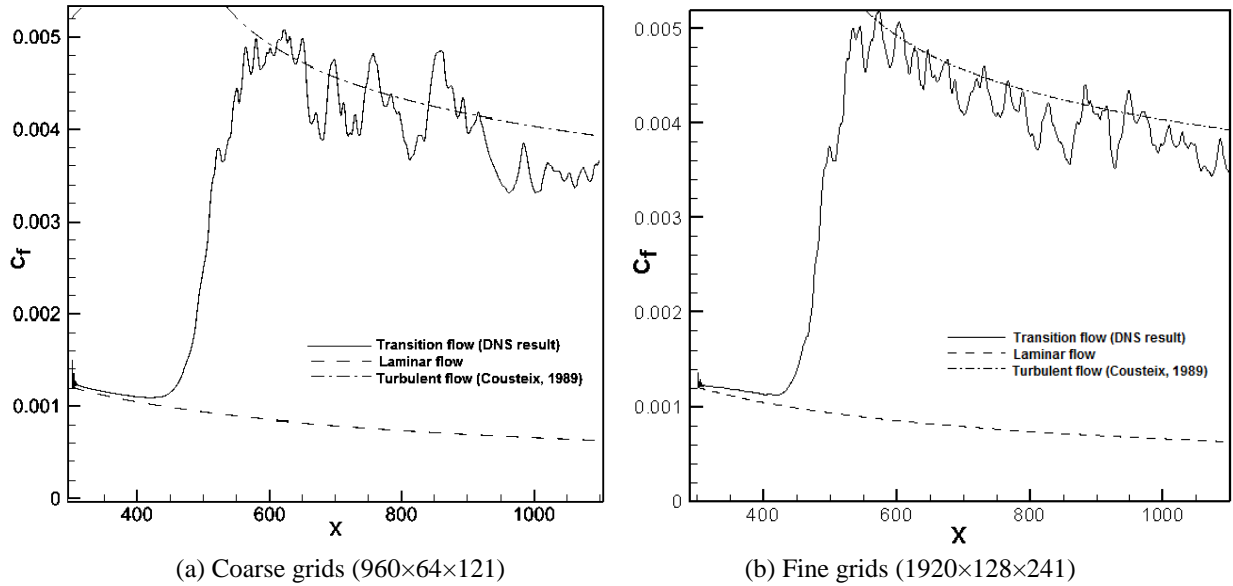

(a) Coarse grids (960×64×121)　　　　　　　　(b) Fine grids (1920×128×241)

**Figure 2 Time-and spanwise-averaged skin-friction coefficient along the streamwise direction**

Figure 3 shows the experiment on the non-linear evolution of the three dimensional soliton-like coherent structure (SCS) in a transitional boundary layer which was proposed by Lee and Li [27]. The results showed that the three-dimensional SCS is the dominant flow structure in almost all dynamic processes in both the early and later stages of boundary layer transitions as well as in a turbulent boundary layer. Figures 2 and 3 have been reported in our previous work. The vortex structures in the transition process of our DNS results are also shown in Figure 4 for comparison. T is a period of T-S wave. It can be seen that the formation and development of hairpin vortices chains of our DNS results is consistent with their experimental work. All these verifications and validations above show that our code is correct and our DNS results are reliable. The Omega identification method (Ω) first proposed by Liu et al. [17] and widely used in our previous studies [13, 28, 29] is also utilized in this paper for capturing the vortex structures. The Ω is defined as the ratio of vorticity squared over the sum of vorticity squared and deformation squared:

$$\Omega = \frac{\|\boldsymbol{B}\|_F^2}{\|\boldsymbol{A}\|_F^2 + \|\boldsymbol{B}\|_F^2}, \quad 0 < \Omega < 1 \qquad (4)$$

where $\boldsymbol{A}$ is the symmetric part of velocity gradient tensor $\nabla \boldsymbol{V}$ and $\boldsymbol{B}$ is the anti-symmetric part, and $\|\cdot\|_F$ is the Frobenius norm. A vortex is identified as the region where $\Omega > 0.5$, which means the vorticity overtakes deformation in this region. This method was also introduced and compared with other vortex identification methods in some review papers and research papers as a new and effective vortex identification method cited by Zhang et al. [30], Epps [31] and Abdel-Raouf et al. [32].

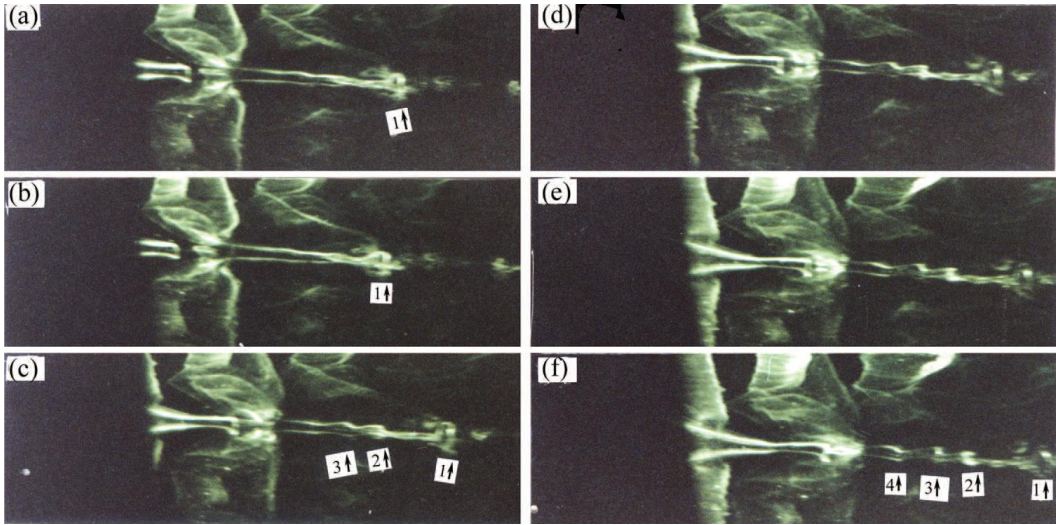

**Figure 3 Evolution of the ring-like vortex chain by experiment [27] at 6 subsequent times (a-f)**

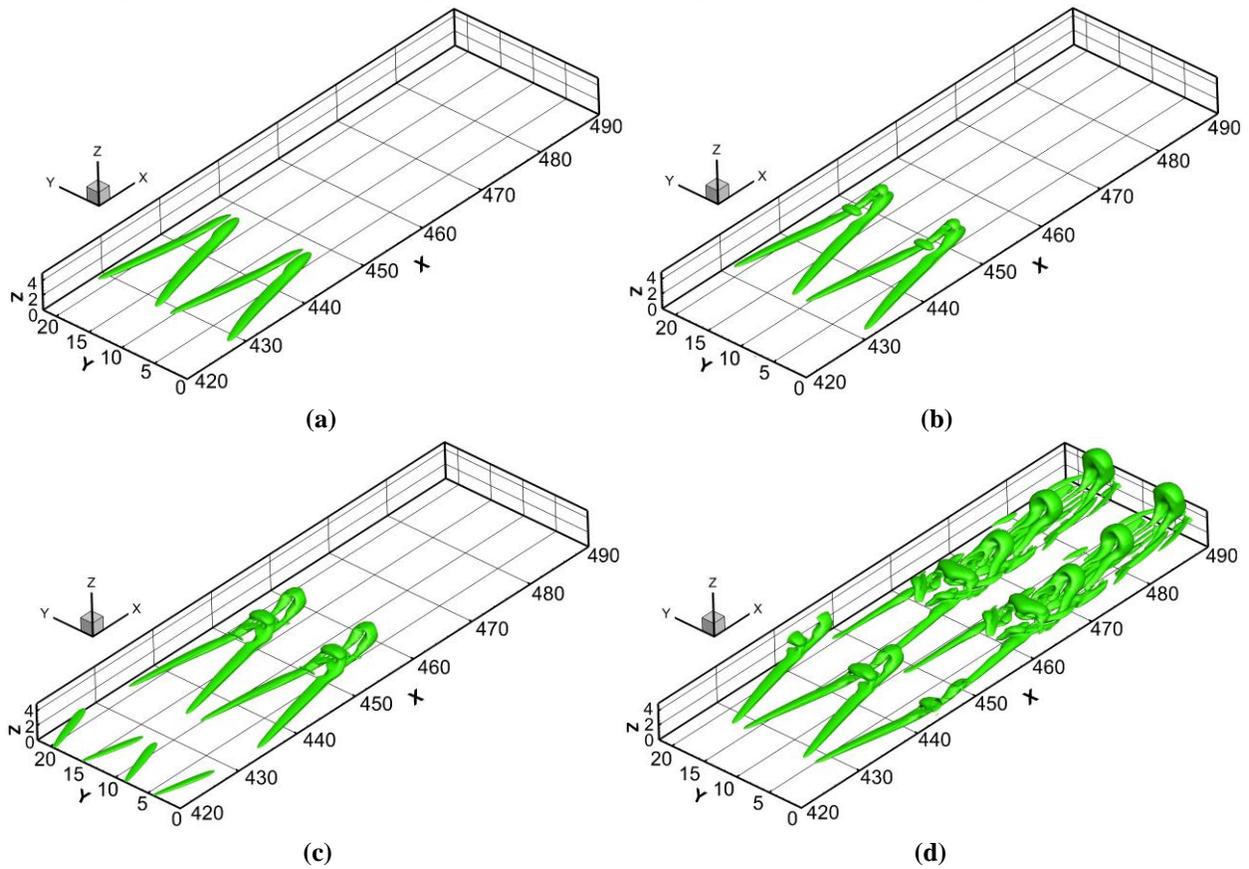

**Figure 4 Vortex structures during the late transition at (a)** $t_1=6.0T$ **(b)** $t_2=6.2T$ **(c)** $t_3=6.4T$ **(d)** $t_4=7.1T$

## 3 Discussion and Results

### *3.1 Vorticity Cannot Represent Vortex*

As proposed by Robinson [**11, 12**], in the turbulent boundary layer, the association can be rather weak between regions of strong vorticity and vortex, especially in the near wall region. For a laminar boundary layer flow with a Blasius solution, the vorticity is concentrated near the wall surface. However, there is no vortex since there is only one-dimensional shear in the boundary layer. It should be noted that when the streamlines are roughly straight, although the magnitude of $\frac{\partial u}{\partial z}$ is large, $\omega_y$ is large as well. In Blasius solution, there is no vortex since the vorticity must overcome the strain rate, which is the reason why $Q > 0$ or $\Omega > 0.5$ is required for the vortex identification. In this section, the comparison between vorticity structures such as vorticity lines or vorticity tubes, and vortex structures like Lambda vortices or hairpin vortices, is given based on our DNS results to reconfirm this viewpoint.

For our case of the flow transition shown in Figure 1, the basic flow is given by a Blasius solution, with the disturbances (T-S waves) being enforced at the inflow boundary. Thus, the transition from laminar flow to turbulence, which gives the vortex structure evolution during the flow transition by iso-surface of $\Omega = 0.501$ (a) at different moments (b) at $t = 15.2T$ (T is a period of T-S wave), is shown in Figure 5. As can be seen, in the early state of transition, although the disturbance is given in the inflow boundary, there is only a shear layer existing near the wall surface without any rotation as both Lambda vortices and hairpin vortices are not generated until the flow develops into the transition state. In the late state of transition, the boundary layer is filled with hairpin vortices. These two types of vortices are confirmed to play a significant role in generation and sustenance of turbulence, and thus the following analyses are based on these two typical vortices.

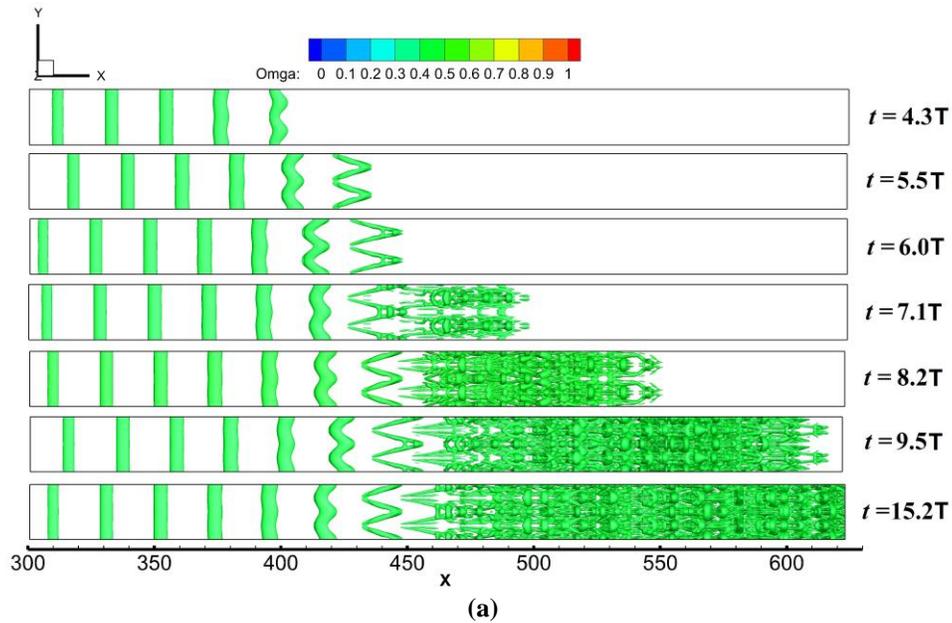

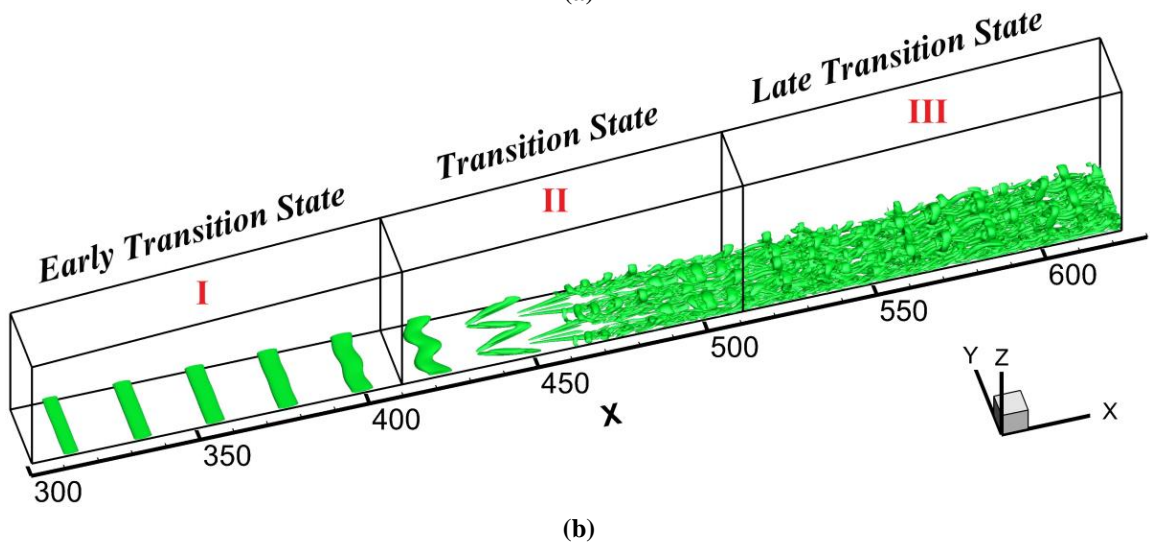

**Figure 5 Vortex evolution during flow transition (a) at different moments (b) three transition states at *t* = 15.2T (iso-surface by Ω method)**

The Lambda vortices and hairpin vortices (by iso-surface of Ω = 0.52) compared with the vorticity lines are respectively shown in Figure 6 and Figure 7. From the vector arrows in Figure 6, all vorticity lines inside the domain start from one side boundary S1 (plane Y= 0 is the side boundary S1, see Figure 1), pass through the Lambda vortices and end at the other side boundary S2. While the hairpin vortices with vorticity lines are shown in Figure 7, as can be seen, these vorticity lines penetrate the leg of the hairpin vortex and, however, better reside in the surface of its head. According to the vector arrows, these vorticity lines also start from one side boundary S1 and end at the other S2. Therefore, it is not proper to regard vorticity lines (tubes) as a vortex to some extent, since a vortex can start or terminate inside the fluid domain but a vorticity tube cannot [4]. A vortex is not always a connected fluid region with a relatively high concentration of vorticity. At the moment when vorticity lines go into the vortex, shown in Figure 6 and Figure 7, the magnitude of vorticity most likely decreases. According to the vorticity flux conservation, the flux area would be larger as vorticity decreases. Therefore, the vortex is not always the congregation of vorticity, but sometimes the dispersion of vorticity, especially for 3-D vortices like the Lambda vortex in a transitional boundary layer. Actually, it is incorrect to determine the vortex structures by vorticity field. Further comparison of the vorticity distribution on the planes which are extracted from the center of Lambda and the hairpin vortex structures is shown in Figure 8. As can be seen from Figure 8(a), the Lambda vortex is located in the region where the vorticity is rather smaller than what is located in its surroundings. From Figure 8(b), an absolute correlation between vorticity congregation and the vortex formation cannot be found. Especially inside the shear layer with high vorticity near the wall, there are no vortex structures.

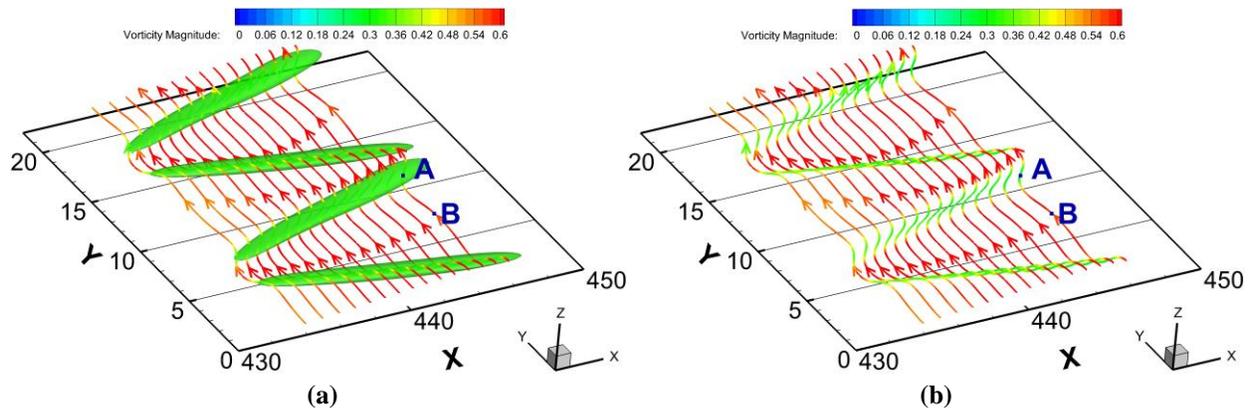

**Figure 6 Lambda vortex and vorticity lines**

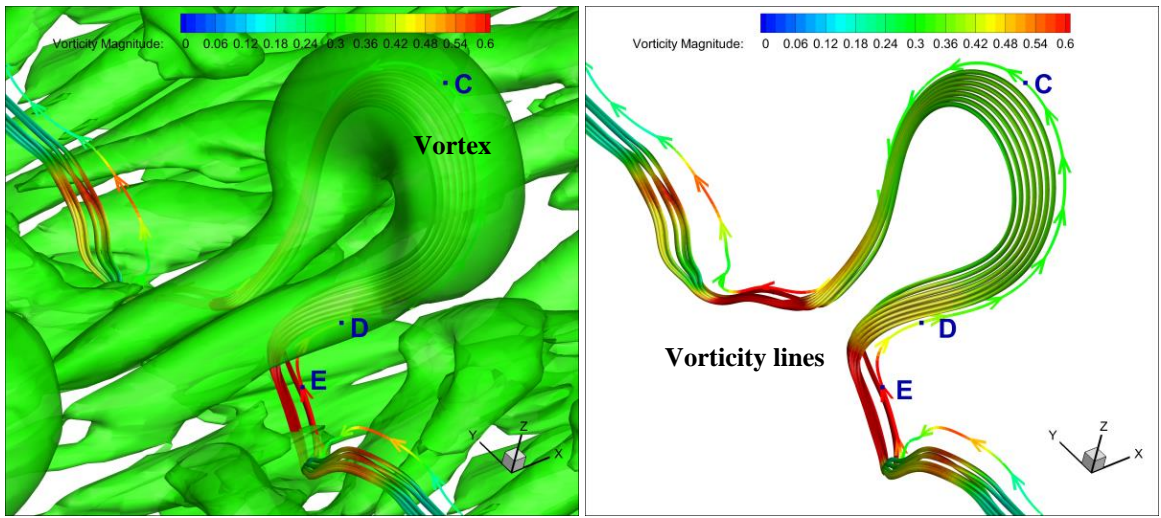

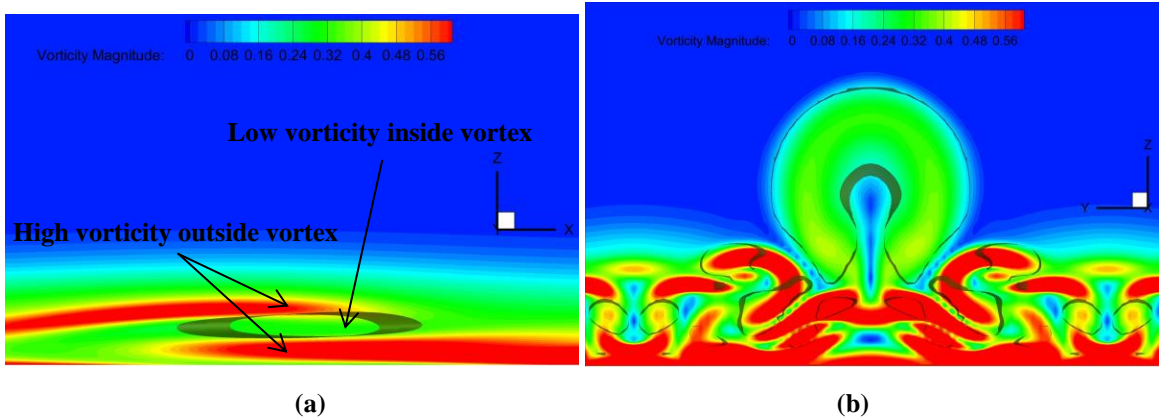

(a)                         (b)

**Figure 8 Vorticity distribution inside the (a) Lambda vortex and (b) the hairpin vortex**

Further quantitative analyses for the comparison of Omega, $\Omega$, and the vorticity magnitude around the region of vortex structures are studied. Two points, A and B, are extracted and detected with values in terms of the $\Omega$ and vorticity magnitude from one single vorticity line, shown in Figure 6. The point A is inside the Lambda vortex surface while the point B is outside the vortex. The detailed values are listed in Table 2. At point A inside the Lambda vortex, the $\Omega$ is about 0.6, which is larger than the value at point B; however, the vorticity is smaller. Three points, C, D, and E are also extracted inside the head and leg of the hairpin vortex, and outside the vortex, shown in Figure 7. From the values of $\Omega$ and vorticity magnitude listed in Table 2, the same conclusion can be obtained. It should be noted that, for one single vorticity line in the whole boundary layer flow domain based on our DNS case, the vorticity magnitude of the vorticity line inside the vortex region is smaller than the value outside the vortex. It is reaffirmed that, the vortex is not absolutely the congregation of vorticity, but is a region where the vorticity overtakes deformation. That means if the deformation is also very large in an area where the vorticity is congregated, it may not be a vortex, since the weight of deformation in the whole velocity gradient is larger than the vorticity, no matter how large the vorticity is. It suggests that the vortex and the vorticity tube are two distinct concepts; the vorticity cannot directly represent a vortex or rotation. That is why the vorticity should be further decomposed into two parts [**18**], rotational vorticity and non-rotational vorticity, readers are encouraged to refer to more details on the vorticity decomposition and a new vortex vector definition proposed by Liu et al. [**33**].

**Table 2 Vorticity and $\Omega$ magnitude inside and outside the Lambda vortices and hairpin vortices**

| Point | Location | $\Omega$ | Vorticity magnitude |
|---|---|---|---|
| A | Inside the Lambda vortex | $\approx 0.60$ | $\approx 0.30$ |
| B | Outside the Lambda vortex | $\approx 0.49$ | $\approx 0.59$ |
| C | Inside the head of the hairpin vortex | $\approx 0.77$ | $\approx 0.31$ |
| D | Inside the leg of the hairpin vortex | $\approx 0.93$ | $\approx 0.47$ |
| E | Outside the hairpin vortex | $\approx 0.43$ | $\approx 0.99$ |

Figure 9 gives the 2D-streamline distribution on the *xoz* central plane near the vortex structures identified by the iso-surface of $\Omega = 0.52$. The background is colored by the vorticity magnitude, and the velocity of free stream is chosen to be the reference frame velocity. It is clearly found that there is a spiral inside the hairpin vortex structures which indicates the motion of rotation; however, the streamlines are not spiral in the region where the vorticity is large (red areas marked with black arrows).

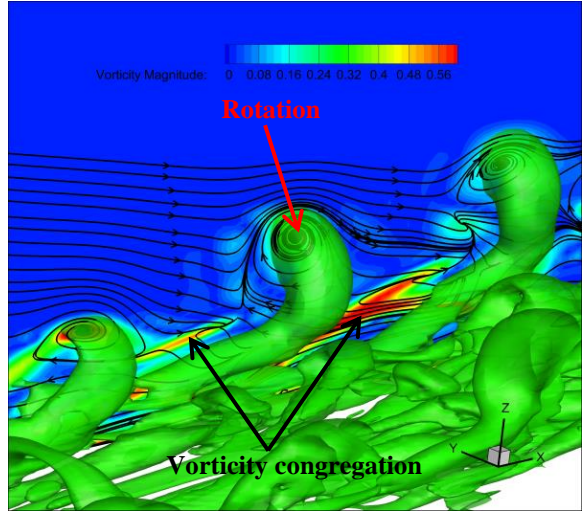

**Figure 9 2D-streamlines distribution around the vortex on central plane**

### 3.2 Vorticity Structures in Flow Transition

#### 3.2.1 Vorticity lines from out boundary

Figure 10 gives the vorticity lines passing through the domain in different states of the flow transition. All of them come from one side boundary and end at the other side boundary, which are obtained from the Blasius solution in the initial flow. Figure 10(a) shows the shortest vorticity tubes in the initial laminar boundary layer, while, vorticity lines in Figure 10(b, c) lift up and are lengthened by stretching, tilting, and twisting.

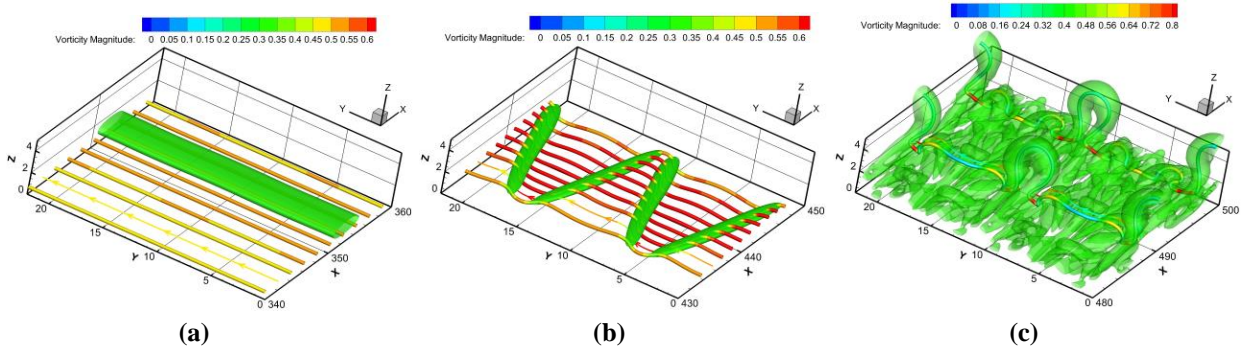

(a)             (b)             (c)

**Figure 10 Vorticity lines in different state of transition**

#### 3.2.2 Vorticity rings

Self-closed vorticity lines, called *vorticity rings*, are generated inside the domain during the flow transition process, and are especially numerous near the hairpin vortex and vortex rings. Figure 11(a, b) show the distribution of vorticity rings outside and inside the hairpin vortices, respectively. Although the vorticity rings have higher vorticity magnitude values inside the hairpin vortex than far outside the vortex structures, the vorticity and vortex structures have very different distributions during transition.

It can be found from Figure 11(a) that two bundles of vorticity rings, marked as 1 and 2 respectively, are located at the upper and lower positions with different vorticity signs. The reason why there are two layers of vorticity rings with different signs generated over the hairpin vortex legs, and which one is generated first, is discussed in the following section.

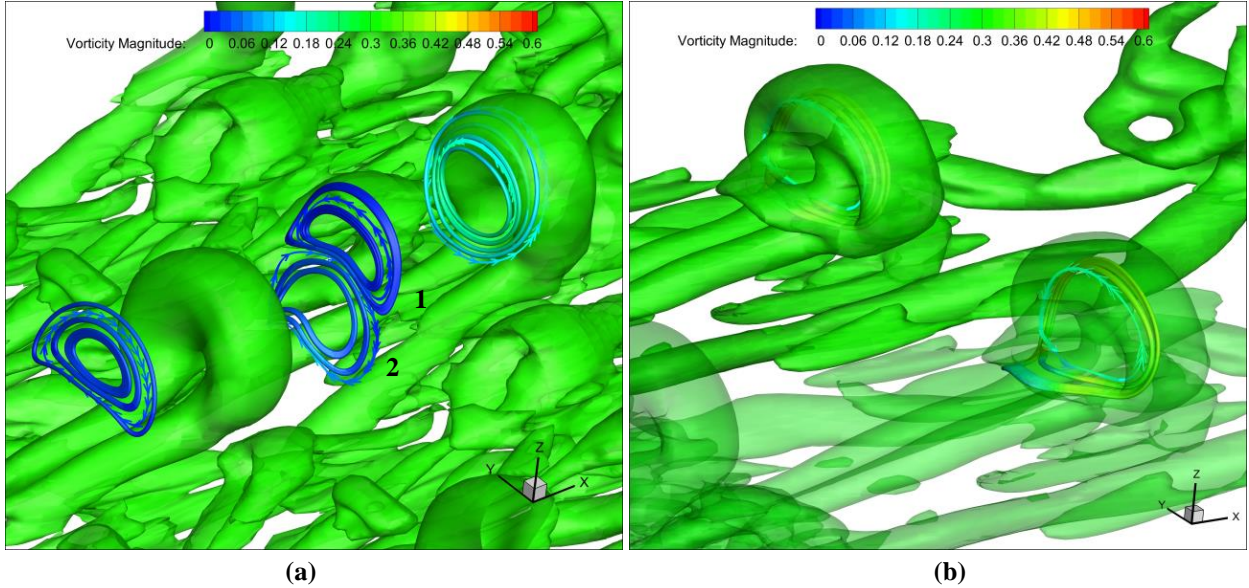

**Figure 11 Distribution of the vorticity rings (a) outside or (b) inside the vortex structures**

*3.2.3 Vorticity structures around/inside vortex structures*

Figure 12 shows the vorticity components inside the Lambda vortex, hairpin vortex, and ring-like vortex. In Figure 12(a), the vorticity lines pass through the legs of the Lambda vortex and display relatively low vorticity magnitude values inside the vortex; for the Lambda vortex, all of vorticity lines are from the outer boundaries. However, in Figure 12(b), two types of vorticity structures can be seen in the hairpin vortex: one type is the vorticity lines coming from the outer boundaries, which enter into one of the vortex legs and go along with the head and then come out through the other leg (marked as 1 in the figure); another is the vorticity rings, part of them are inside the vortex and part of them are outside (marked as 2). For the ring-like vortex shown in Figure 12(c), there are three components of vorticity, two are the same as 1, 2 shown in Figure 12(b), and the third part is the vorticity rings located totally inside the vortex structures (marked as 3 in Figure 12(c)). Therefore, it is not easy to detect the vortex structures, especially the legs of vortex by vorticity structures, which is why we emphasize that the vorticity cannot be a useful signal of flow transition from the laminar flow to turbulence flow.

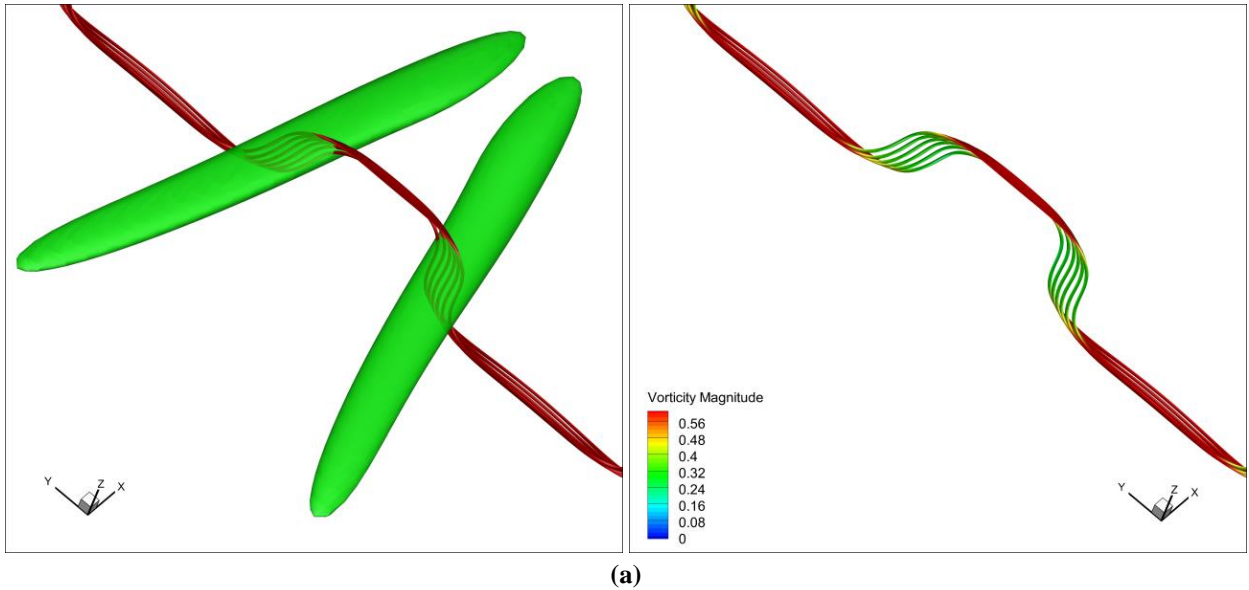

(a)

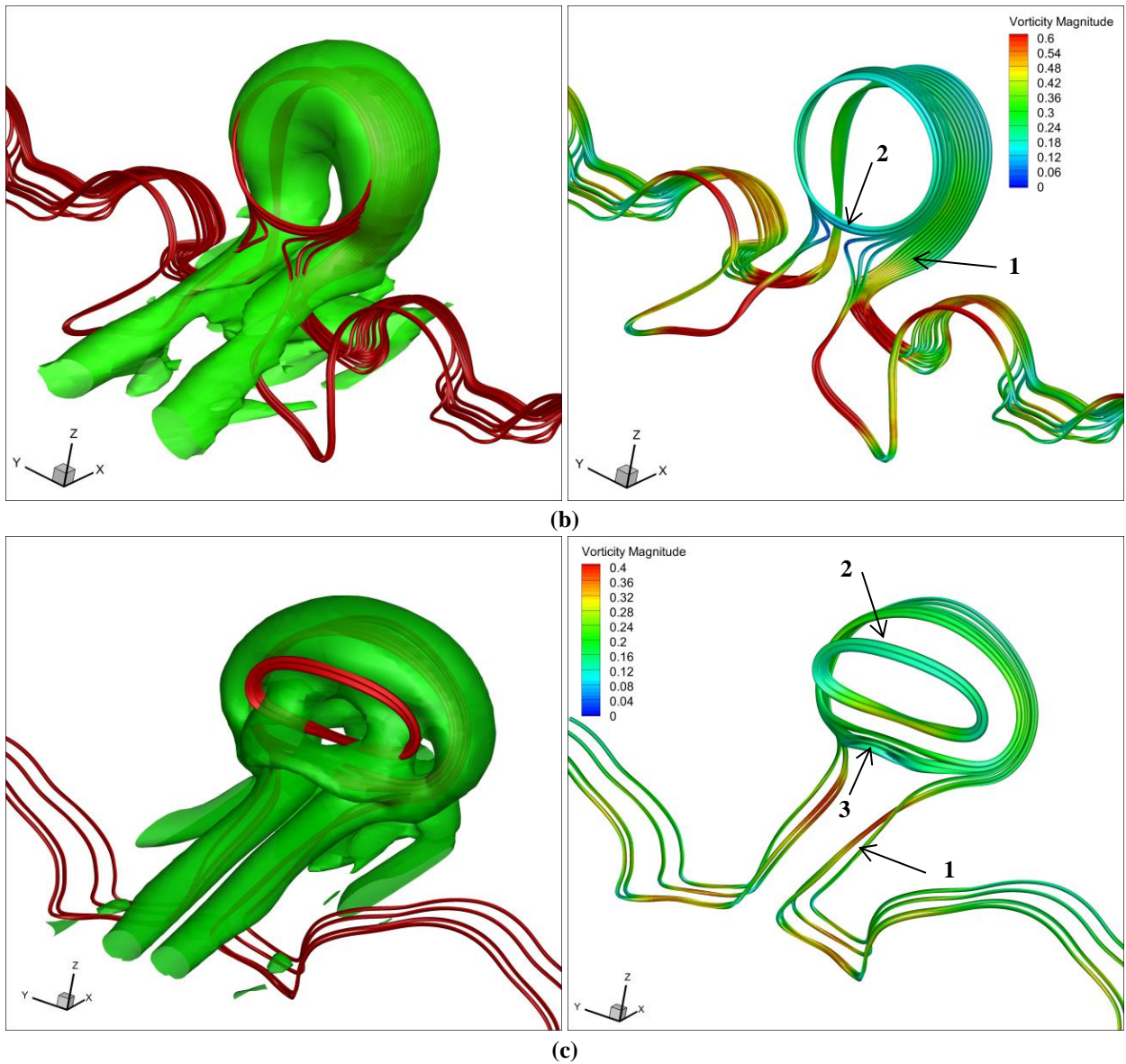

**Figure 12 Vorticity components inside (a) Lambda vortex (b) hairpin vortex (c) ring-like vortex**

### 3.3 Formation of Vorticity Rings

*3.3.1 Vorticity rings generated in early transition*

The vorticity rings were generated in a very early transition state, and they can take different shapes during the generation process. Here, we only focus on the spanwise- and normal-dominant vorticity rings in the early state of flow transition since the vorticity structures at this stage are more organized. Figure 13 shows the vorticity line distribution on the planes extracted from X=301 to 307 in the spanwise direction. Figure 13(a) plots the vorticity lines at the beginning of the flow, which are all from the side boundary S1 (plane Y=0) to S2 (plane Y=22). However, due to the disturbance, the vorticity lines near the wall boundary become more and more crooked (see Figure 13(b)), until the vorticity rings are generated where a converse vorticity gradient appears. From Figure 13(c), the vorticity ring appears in the bottom first. Then, in Figure 13(d, e), another two vorticity rings are generated in the second layer and both have the opposite signs to the previous one. So far two layers of vorticity rings are formed in Figure 13(f).

It should be noted that the vorticity rings are self-closed and newly generated inside the domain. However, it must be explained why these vorticity rings are generated and increase inside the domain. The vorticity transport equation of fluid dynamics in the case of incompressible (low Mach number) and Newtonian fluids can be written as,

$$\frac{D\vec{\omega}}{Dt} = \frac{\partial \vec{\omega}}{\partial t} + (\vec{V}\cdot\nabla)\vec{\omega} = (\vec{\omega}\cdot\nabla)\vec{V} + \nu\nabla^2\vec{\omega} \tag{5}$$

where $\vec{V}$ is the flow velocity, $\nu$ is the kinematic viscosity and $\nabla^2$ is the Laplace operator. The term $(\vec{\omega}\cdot\nabla)\vec{V}$ is the gradient of the velocity $\vec{V}$ projection in the direction of the vorticity vector multiplied by the vorticity magnitude. For a three-dimensional viscous flow, the vorticity on the right-hand side of Eq.5 can only change for two reasons based on these two terms: $(\vec{\omega}\cdot\nabla)\vec{V}$ and $\nu\nabla^2\vec{\omega}$. On the one hand, according to the term, $\nu\nabla^2\vec{\omega}$, vorticity can change through diffusion in the flow due to the action of viscosity; On the other hand, new vorticity can be generated because of the term, $(\vec{\omega}\cdot\nabla)\vec{V}$, which is attributed to the stretching, tilting or twisting of the vorticity tubes. However, according to Helmholtz vorticity conservation law, the vorticity lines cannot interrupt, start, or end inside the domain but can only have one form without exception – vorticity rings. This clearly leads to a conclusion that the vortex generation, due to vorticity stretching, tilting or twisting, can produce new vorticity which can only have a unique form, vorticity rings only, since the vorticity lines cannot be interrupted or ended inside the flow field.

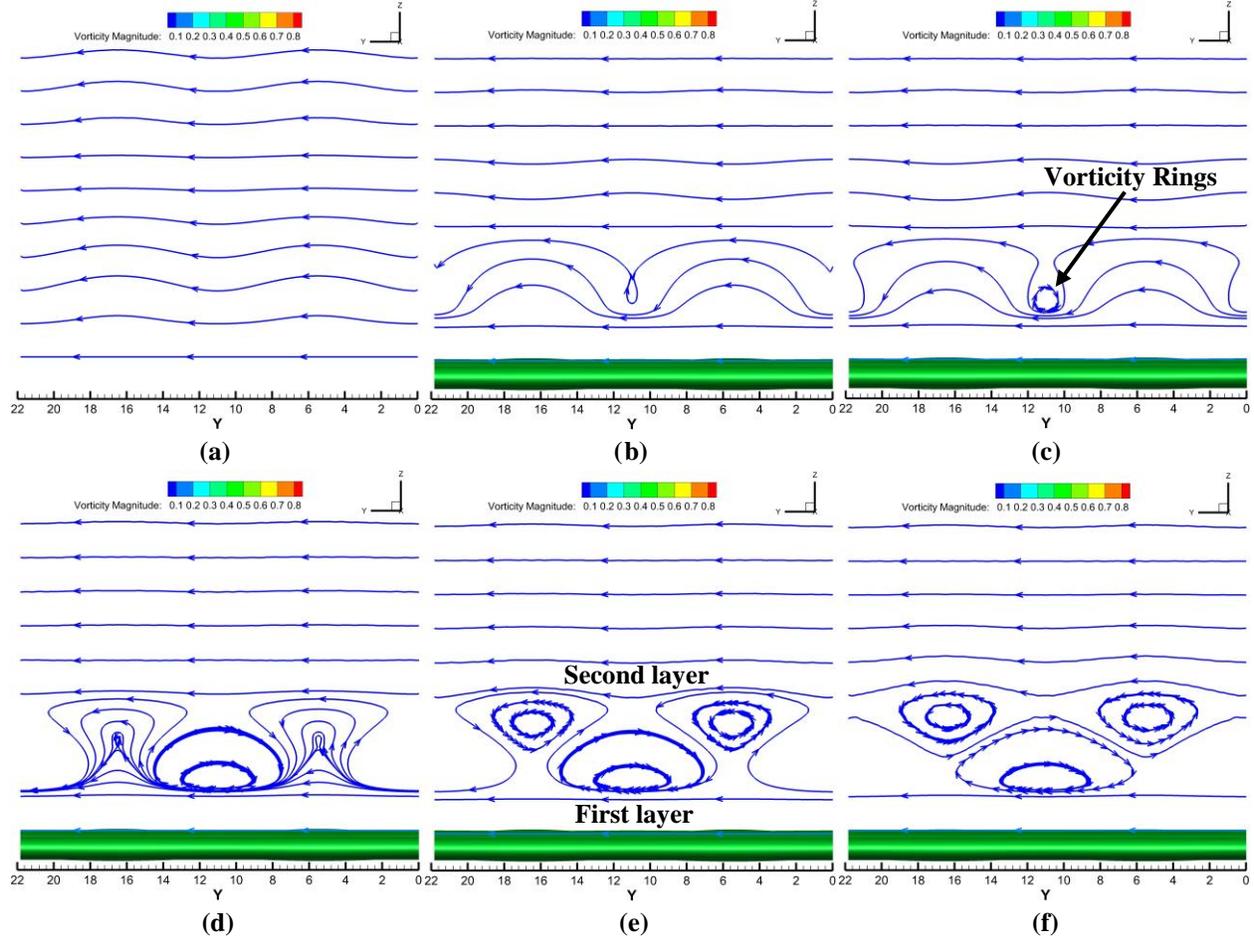

Figure 13. Formation of the vorticity rings in early transition

*3.3.2 Vorticity rings generated near hairpin vortex*

Although it is confirmed that vorticity cannot represent the vortex, and also a vortex is not absolutely a region where the vorticity is congregated, vorticity rings shown in Figure 11 have higher values of vorticity magnitude when they are closer to the vortex, especially inside the vortex. A detailed comparison is shown in Figure 14. The vorticity magnitude values of the vorticity rings far away from the vortex structures are very small, approximately 0.01, while the vorticity rings closer to the vortex, especially inside the vortex, have much higher values of vorticity magnitude, around 0.4. As mentioned above, the newly generated vorticity is attributed to the stretching, tilting or twisting of the vorticity lines. In fact, the generation of vortex structures is caused by the lengthening (stretching, tilting or twisting) of vorticity lines. During the vortex generation process, the vortex rings generate a momentum deficit which causes high shear layers around them. These high shear layers are new vorticity. However, all newly generated vorticity can only have a unique form without exception – vorticity rings. Therefore, we conclude that the vorticity rings are mainly generated and developed along with the vortex, which means the new vorticity rings' generation and growth are attributed to the buildup of the vortex structure. However, it should be pointed out that the newly generated vorticity rings still cannot be called vortices, since the vorticity rings appear in the region where there is no vortex, and the vorticity magnitude is relatively very small.

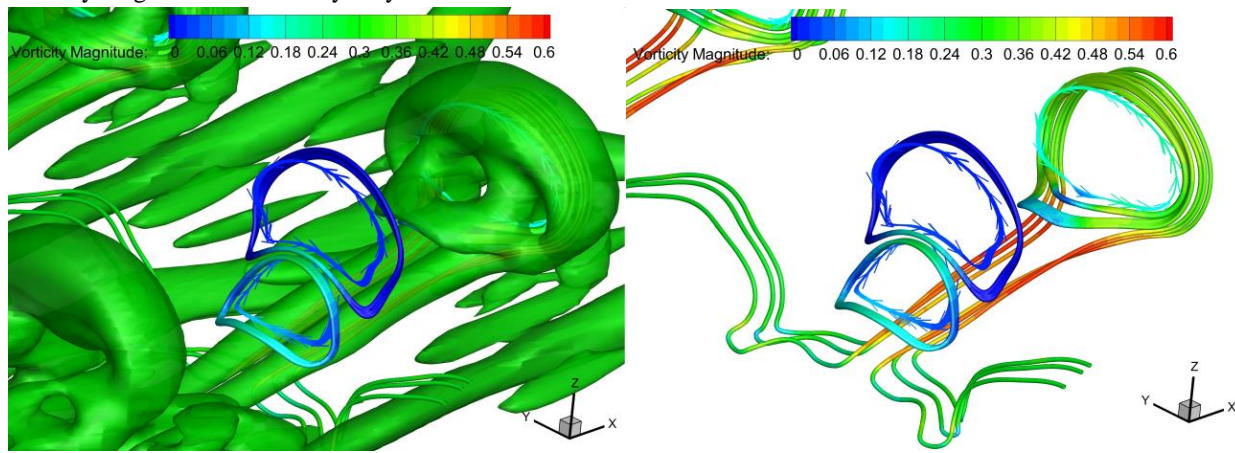

**Figure 14. Comparison of the vorticity magnitude between the vorticity rings outside / inside the vortex structures**

## 4 Conclusions

In this paper, different types of vorticity structures (not vortex structures) in the process of laminar boundary layer transition flow on a flat plate, are compared with the vortex structures based on our high-order direct numerical simulation (DNS). Several conclusions obtained from our DNS results are described as follows:

(1) The vorticity cannot directly represent vortex or rotation. A vortex is not absolutely the congregation of vorticity, but is a region where the vorticity overtakes deformation, since the weight of deformation in the whole velocity gradient cannot be ignored, no matter how large the vorticity is.

(2) Except for the vorticity lines which come from and end at side boundaries, another type of vorticity structure, self-closed vorticity lines, which can be called *vorticity rings*, are numerously generated inside the domain during the transition process. There are more layers of vorticity rings generated around vortex structures.

(3) The generation and growth of vorticity rings are attributed to the buildup of the flow vortex structure. The basic reason for their generation is that the new vortex rings generate a low speed zone which creates a high shear layer around them. The high shear layer is new vorticity near the vortex rings. However, all newly generated vorticity can only have a unique form, rings, without exception, since the vorticity lines cannot interrupt, start, or end inside the flow field. We believe this mechanism describes why we can find many vorticity rings near vortex rings. Vorticity rings in other places should have a similar mechanism of generation.

(4) Vorticity and vortex are different but very closely related. According to our DNS, vortex is generated by vorticity line stretching, tilting and twisting, but the new vorticity rings are generated by the vortex buildup. When the vortex is formed, a high shear will be generated duo to the vortex rotation or so-called ejections and sweeps. These new shear layers are the source of new vorticity generation.

(5) Vortex consists of variety of vorticity lines in flow transition. In Lambda vortex, almost all of vorticity lines come from and end in side boundaries. In the hairpin vortex, the vorticity lines mainly come from side boundaries with some of exceptions which are vorticity rings. In the vortex rings, most of the vorticity lines are self-closed, some of them come from the side boundaries and some of them are part of vorticity rings.

## Acknowledgments

This work was supported by Department of Mathematics at University of Texas at Arlington. The authors are grateful to Texas Advanced Computing Center (TACC) for the computation hours provided. This work is accomplished by using Code DNSUTA released by Dr. Chaoqun Liu at University of Texas at Arlington in 2009. Xiangrui Dong also would like to acknowledge the China Scholarship Council (CSC) for financial support.